\documentclass[aps,prl,twocolumn,groupedaddress,nofootinbib,notitlepage,showpacs,floatfix]{revtex4-1}

\usepackage{graphicx,graphics,epsfig,subfigure,times,bm,bbm,amssymb,amsmath,amsfonts,amsthm,mathrsfs,MnSymbol}
\usepackage[matrix,frame,arrow]{xypic}
\usepackage[pdfstartview=FitH]{hyperref}
\usepackage{pifont}
\hypersetup{
    colorlinks=true,       
    linkcolor=red,          
   citecolor=magenta,        
    filecolor=magenta,      
    urlcolor=cyan,           
    runcolor=cyan
}
\usepackage[pdftex]{color}
\usepackage{braket}
\usepackage{enumerate}
\usepackage[normalem]{ulem}

\usepackage{subfigure}
\usepackage{overpic}

\usepackage{multirow}
\newcommand{\be}{\begin{equation}}
\newcommand{\ee}{\end{equation}}
\newcommand{\ba}{\begin{eqnarray}}
\newcommand{\ea}{\end{eqnarray}}
\newcommand\tr{{\mbox{Tr\,}}}
\newcommand{\ignore}[1]{}

\begin{document}
\title{
Local response of  topological order to an external perturbation
}
\author{Alioscia Hamma}
\affiliation{Center for Quantum Information,
Institute for Interdisciplinary Information Sciences,
Tsinghua University, Beijing 100084, P.R. China}
\affiliation{Perimeter
Institute for Theoretical Physics, 31 Caroline St. N, N2L 2Y5,
Waterloo ON, Canada}
\author{Lukasz Cincio}
\affiliation{Perimeter
Institute for Theoretical Physics, 31 Caroline St. N, N2L 2Y5,
Waterloo ON, Canada}
\author{Siddhartha Santra}
\affiliation{Department of Physics and Astronomy \& Center for Quantum Information Science and Technology,University of Southern California, Los Angeles, California 90089-0484, USA}
\author{Paolo Zanardi}
\affiliation{Department of Physics and Astronomy \& Center for Quantum Information Science and Technology,University of Southern California, Los Angeles, California 90089-0484, USA}

\author{Luigi Amico}
\affiliation{CNR-MATIS-IMM $\&$ Dipartimento di Fisica e Astronomia Universit\`a di Catania, C/O ed. 10, viale A. Doria 6,
  95125 Catania, Italy}
  \affiliation{Perimeter
Institute for Theoretical Physics, 31 Caroline St. N, N2L 2Y5,
Waterloo ON, Canada}

\begin{abstract}
We study the behaviour of the R\'enyi entropies for the toric code subject to a variety of different perturbations, by means of 2D DMRG and analytical methods. We find that R\'enyi entropies of different index $\alpha$ display derivatives with opposite sign,  as opposed to typical symmetry breaking states, and can be detected on a very small subsystem regardless of the correlation length. This phenomenon is due to the presence in the phase of a point with flat entanglement spectrum, zero correlation length, and area law for the entanglement entropy. We argue that this kind of splitting is common to all the phases with a certain group theoretic structure, including quantum double models,  cluster states, 
and other quantum spin liquids. The fact that the size of the subsystem does not need to scale with the correlation length, makes it possible for this effect to be accessed experimentally.
\end{abstract}
\maketitle

{The understanding of collective behaviour  arising from  microscopic  interactions is one of the central issues in physics. Indeed, the most recent research has been indicating that the kind of emerging order in extended systems is not so simple as  was  expected. This is the case of quantum phases of matter with the so-called topological order. These phases, and the phase transitions between them, cannot  be characterized within the symmetry breaking mechanism, a cornerstone in  many-body physics. 
Topological phases  are characterized by global correlations that are  captured by subtle entanglement properties.
This  makes  topological order an elusive kind of order, and therefore  very difficult to detect. 
More generally, quantum spin liquids cannot be characterized by a local order parameter, and are therefore very difficult to detect. 
Here we show  that the order in a paradigmatic class of  spin liquids, some of which are topological,
can be revealed by measurements  on a small portion of the system through the analysis of their  response  to an external perturbation.  
To this aim, we employ    analytical and  numerical methods  in   quantum many-body theory and quantum information. 
Our results can provide a venue to detect topological order experimentally.}

The Landau symmetry breaking mechanism classifies different phases of many-body systems according to the symmetry  that
the system-low-energy-states breaks
\cite{Goldenfeld}. 
This implies that some observable  exists whose values tell us in what phase the system is, the so called order parameter. In this picture, the  macroscopic  order arises from  local order.   For instance, we can distinguish a magnet from a paramagnet by measuring the magnetization resulting from the breaking of a  local rotation symmetry of the atomic spin. 
 
 In the last two decades,  a more complex scenario emerged. An example of paramount importance in  condensed matter physics is provided by  quantum phase transitions  occurring in two dimensional electron gas displaying  fractional quantum Hall effect\cite{Fractional-Hall} : Different Hall states  are characterized by different physical properties,  despite sharing the same symmetries. Different types of topological orders have been found in various physical systems so far,  like spin liquids \cite{White_liquids}, anyonic systems\cite{Wen-1}, topological insulators\cite{topo_ins}. Since these states do not break any symmetry, they cannot be distinguished by measuring a local order parameter.  The comprehension of topological phases is one of the central issues in  modern condensed matter\cite{Wen-1}. Topological states have  properties like protected edge modes or degeneracies that are 
robust under arbitrary perturbations\cite{Wen-1}.  They possess anyonic excitations whose interactions are of topological nature and are inherently robust against decoherence. These features have made such states very interesting  as a platform for quantum computers\cite{tqc}.

As a result from the interaction between condensed matter and quantum information, it has been  understood   
that  topological order in the ground state of a given system can be characterized through entanglement\cite{Amico_rev,Cramer_rev, te}. { Specifically, a long range pattern of entanglement  results in the ground state, called topological entanglement entropy} $\gamma$\cite{te}.  Such a quantity can be computed    through the von Neumann entropy $S_1$ of the reduced density matrix  $\rho_A$ describing the subsystem $A$
(being the system $A\cup B$ in  pure state). 
 For a regular region $A$ then von Neumann entropy is given by { $S_1=  a |\partial A| - \gamma +g$}, where $|\partial A|$ measures the boundary of $A$. 
The quantities $a,g$ are non universal and depend on the details of the Hamiltonian describing the system, like the correlation length $\xi$. The finite correction $\gamma$, instead, is universal and can serve as partial classification of topological phases\cite{renyi, num}.  
We remark that, although $A$ is a finite region, $S_1$ is not a local order parameter, because it is not the expectation value of a local observable.  
On the other hand, the  reconstruction of $\rho_A$ by local measurements  seems to be quite a daunting task because  $A$, though finite, scales with $\xi$\cite{te}. Larger system sizes can be studied numerically by using R\'enyi entropies through the replica trick\cite{Isakov}.

\begin{figure}[t!]
	\centering
\includegraphics[width=0.7\columnwidth,clip=true]{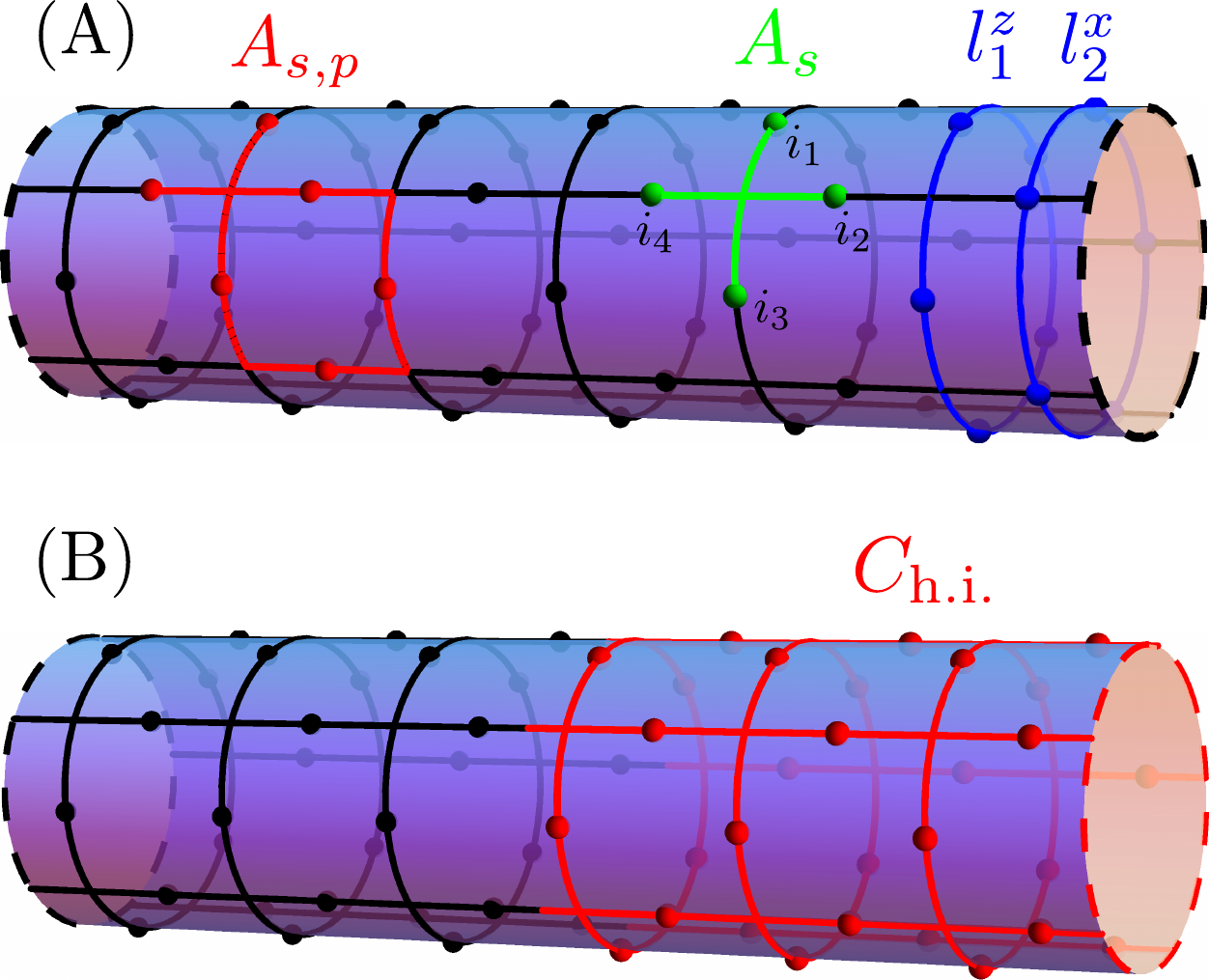}
	\caption {
	Cylinder of infinite length and width $L_y = 5$ used in 2D DMRG calculation. (A) Subsystems on which R\'enyi entropies are  calculated: $A_s$ - one star and $A_{s,p}$ - composition of star and plaquette. Loops $l^z_1$ and $l^x_2$ used to distinguished between topological sectors are also depicted. (B) Subsystem $C_\textrm{h.i.}$ that contains half of the infinite cylinder .
}
	\label {cylinder} 
\end {figure}

Much effort has been devoted to analyze the topological order in the ground state for a fixed value of the control parameter $\lambda$. In this paper, we push forward the idea that  progress in probing topological order can be achieved  by analyzing  {\it local  variations} of the ground state by external perturbation.

We study a set of spin$-1/2$ localized at the edges of  $2D$ square lattice with periodic boundary conditions in presence of a perturbation $V$: \begin{equation}
\mathcal{H} = - \sum_s  \prod_{i \in s} \sigma^x_i - \sum_p \prod_{i \in p} \sigma^z_i + V(\lambda)
\label{model}
\end{equation}
where $s$ and $p$ label the vertices and plaquettes of the lattice respectively, while   $\sigma^x_i, \sigma^z_i$ are  Pauli operators of the spin living at  the edge $i$. 
For $V(\lambda)=0$  the Hamiltonian above is the celebrated toric code, a paradigmatic model for topological order\cite{Kitaev}. For the analysis below, we remark that in this case, the ground 
state of this model features  $\xi=0$. We consider different  $V(\lambda)$  (see Table \ref{table}) where $\lambda$ stands for $\{ \lambda_1,\dots,\lambda_N\}$ that are the  parameters controlling the perturbation.  The perturbation in (\ref{model}) such that the correlation length is increasing with $\lambda_i$ until divergence at the critical point $\lambda_c$. For a discussion of the critical point, see \cite{critical}.

 For $\lambda<\lambda_c$ these systems are topologically ordered, while for $\lambda>\lambda_c$ they are trivial paramagnets. In both phases there is no local order parameter. This model belongs to a class of so-called quantum double models which correspond to those phases whose low energy theory is a lattice gauge theory \cite{Kitaev}.
 In this article, we demonstrate how the topological phase can be distinguished from the   paramagnet of (\ref{model}) through the inspection of
 \ignore{
We   demonstrate that   one can distinguish the topological from the paramagnetic phases of (\ref{model}) through the inspection of the response of a small subsystem to the external perturbation $V(\lambda)$.
 In particular, we will show that the topological phases of (\ref{model}) are characterized by a specific behaviour} $\partial_\lambda S_\alpha$, where $S_\alpha \doteq (1-\alpha)^{-1} \log \tr \rho_A^\alpha$ are the R\'enyi entropies. We shall see that  $\partial_\lambda S_\alpha$ present a peculiar behavior even for a small subsystem $A$ as $\xi$ varies as an effect of a perturbation.
We remind  that $S_0 =\log R$, where $R$  is the Schmidt rank of the state, i.e. the number of nonzero eigenvalues of $\rho_A$, and that $S_1$ is the von Neumann entropy measuring the entanglement entropy for the subsystem $A$.  We observe that, generically,   $R$ and thus $S_0$  increase with $\xi$  because more degrees of freedom get entangled by increasing the  room for correlations.
On the other hand, we find that the behaviour of $S_1$ is drastically different between topological phases and the topologically trivial phases presented here. 

\begin{figure}[t!]
	\centering
\includegraphics[width=0.9\columnwidth,clip=true]{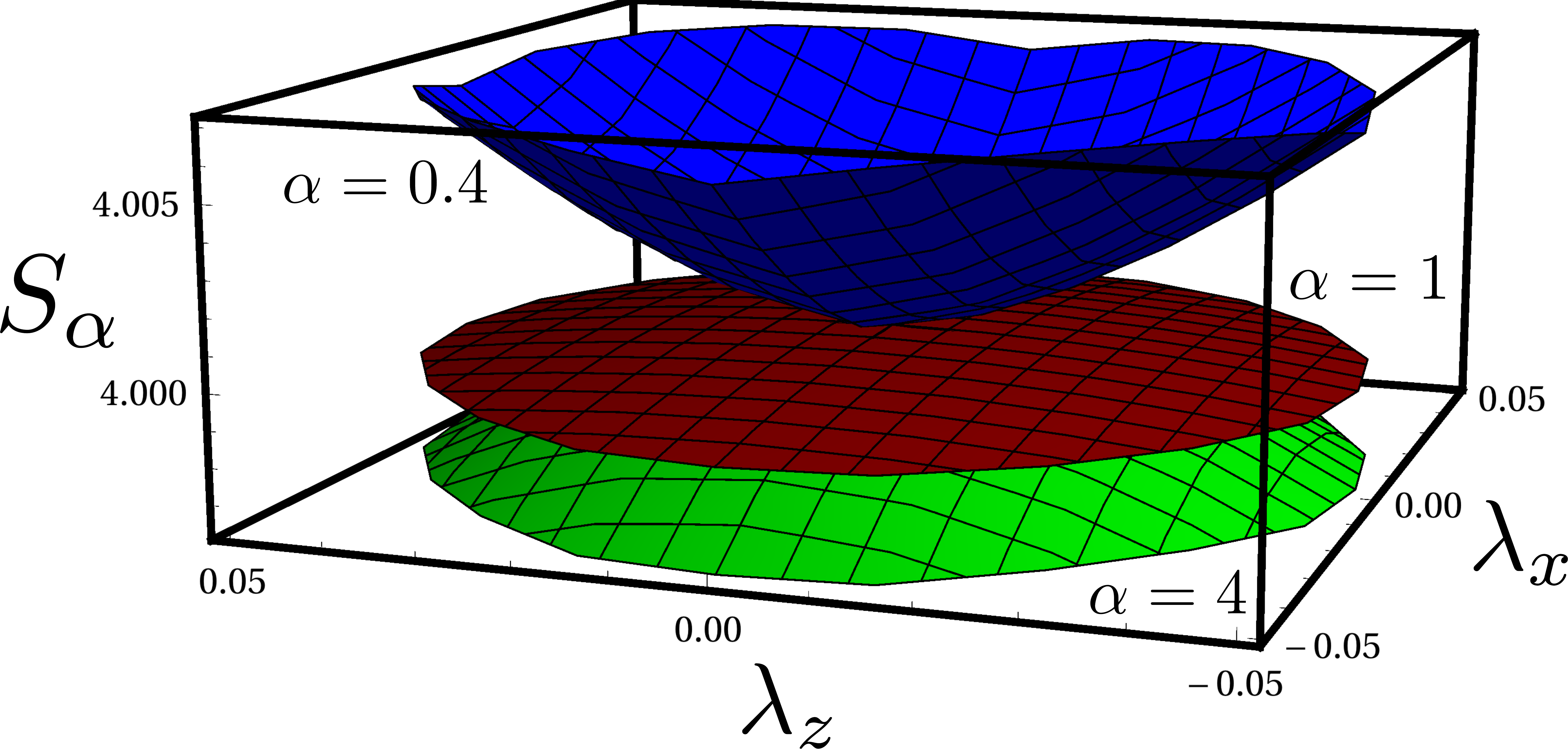}
	\caption {The splitting phenomenon. The figure displays the splitting with opposite slopes between the small and large $\alpha$ R\'enyi entropies. { We see the splitting occurring around $\alpha \simeq 0.6$.} The R\'enyi entropies are calculated for the partition $A_{s,p}$ of Fig.\ref{cylinder}A  for the ground state  of $\mathcal H= \mathcal H_{TC} + V_{xz}$. }
	\label {3dplot} 
\end {figure}

Our approach elaborates on the quantum information notion of  differential local convertibility (DLC), which states that two {proximal} bipartite pure states are locally convertible if it is possible to transform one into the other by resorting only on local quantum operations on $A$ and $B$ separately, plus classical communication and the aid of an ancillary entangled system\cite{Plenio_lc}. The necessary and sufficient condition for that is {$\partial_\lambda S_\alpha < 0$ for all $\alpha > 0$} \cite{Turgut_lc}. 
In  Ref.\cite{Cui_locc} such notion has been  applied to the order-disorder quantum phase transition occurred in a one dimensional spin system.

For each $V(\lambda)$ we compute the ground state wavefunction $|\psi(\lambda)\rangle$ and its reduced density matrix $\rho_A (\lambda)$. For some $V(\lambda)$ 
we can apply  exact analytical approach; for  the  generic perturbation $V_{xz}(\lambda)= \sum_i (\lambda_z {\sigma^z_i}+\lambda_x {\sigma^x_i})$ we  resort to numerical analysis.  
\begin{table}[height=3cm]
\centering  
\begin{tabular}{c c c c c} 
\hline\hline                        
Perturbation $V(\lambda)$ & G.I. & DLC & Exact & $\xi$ \\ 
\hline                  
$\sum_se^{-\lambda_s\sum_{i\in s}{\sigma^z_i}}$ & \ding{51} & \ding{51} &  \ding{51} & 0 \\ 
$\lambda_h\sum_{i\in H}{\sigma^z_i}$ & \ding{51} & \ding{55} &  \ding{51} & $\ne 0$ \\
$\lambda_z \sum_{i}{\sigma^z_i}$ & \ding{51} &  \ding{55} & \ding{55} &$\ne 0$\\
$\lambda_z\sum_{i}{\sigma^z_i}+\lambda_x\sum_{j}{\sigma^x_j}$ & \ding{55} & \ding{55}   & \ding{55} & $\ne 0$ \\ [1ex]      
\hline 
\end{tabular}
\caption{Local Convertibility in Topological Phase. DLC, i.e no splitting of the R\'enyi entropies, only occurs if the perturbation is fine tuned in order to keep the system with $\xi=0$.
The left column shows the type of perturbation studied. The first column details whether the considered model is Gauge Invariant. The second column  indicates whether  DLC occurs. For certain perturbations the ground state of the system is accessible exactly (third column). The last column provides the information on $\xi$. The subsystems we refer here  have a non trivial bulk\cite{long} .}
\label{table}
\end{table}

The numerical method employed here is an infinite DMRG algorithm \cite{DMRG} in two dimensions, detailed in \cite{Cin12b}. The method provides Matrix Product State (MPS) representation of a complete set of ground states on a cylinder of infinite length and finite width $L_y$ (Fig.\ref{cylinder})
for a given Hamiltonian that realizes topological order. As argued in \cite{Cin12a}, each ground state has a well-defined flux threading through the cylinder. The flux is measured by (in general) dressed Wilson loop operators that enclose the cylinder in the vertical direction.

In the case of fixed point toric code (Eq.\ref{model} with $V=0$), these loops are given by $l^z_1$ and $l^x_2$ (Fig.\ref{cylinder}A). Four topological sectors are then distinguished by $\langle l^z_1 \rangle, \langle l^x_2 \rangle = \pm 1$. Once the perturbation is present, Wilson loops may change, but as long as the perturbation is small, $\langle l^z_1 \rangle$ and $ \langle l^x_2 \rangle$ can still be used to identify topological sectors because $\langle l^z_1 \rangle, \langle l^x_2 \rangle \simeq \pm 1$.

Simulations are carried out with cylinders of width up to $L_y = 5$ for $\sqrt{\lambda_x^2 + \lambda_z^2} \leq 0.05 $ {and $0 \leq \lambda < 0.7$ as shown in Fig.\ref{3dplot} and Fig.\ref{p2} respectively}. {In the topological phase,  the} outcome of each simulation is four quasi-degenerate ground states, from which the one with $\langle l^z_1 \rangle, \langle l^x_2 \rangle \simeq + 1$ is chosen for further investigation. {This is done to ensure that finite size effects have the least possible impact on results. In the limit $L_y \rightarrow \infty$ all four ground states become locally indistinguishable.  The results are converged in bond dimension of MPS which acts as a refinement parameter. A reduced density matrix of a half-infinite cylinder $C_\textrm{h.i.}$ (Fig.\ref{cylinder}B) is calculated throughout the simulation. The bond dimension is increased until convergence of its spectrum is reached. 


\begin{figure}[t!]
	\centering
\includegraphics[width=0.9\columnwidth,clip=true]{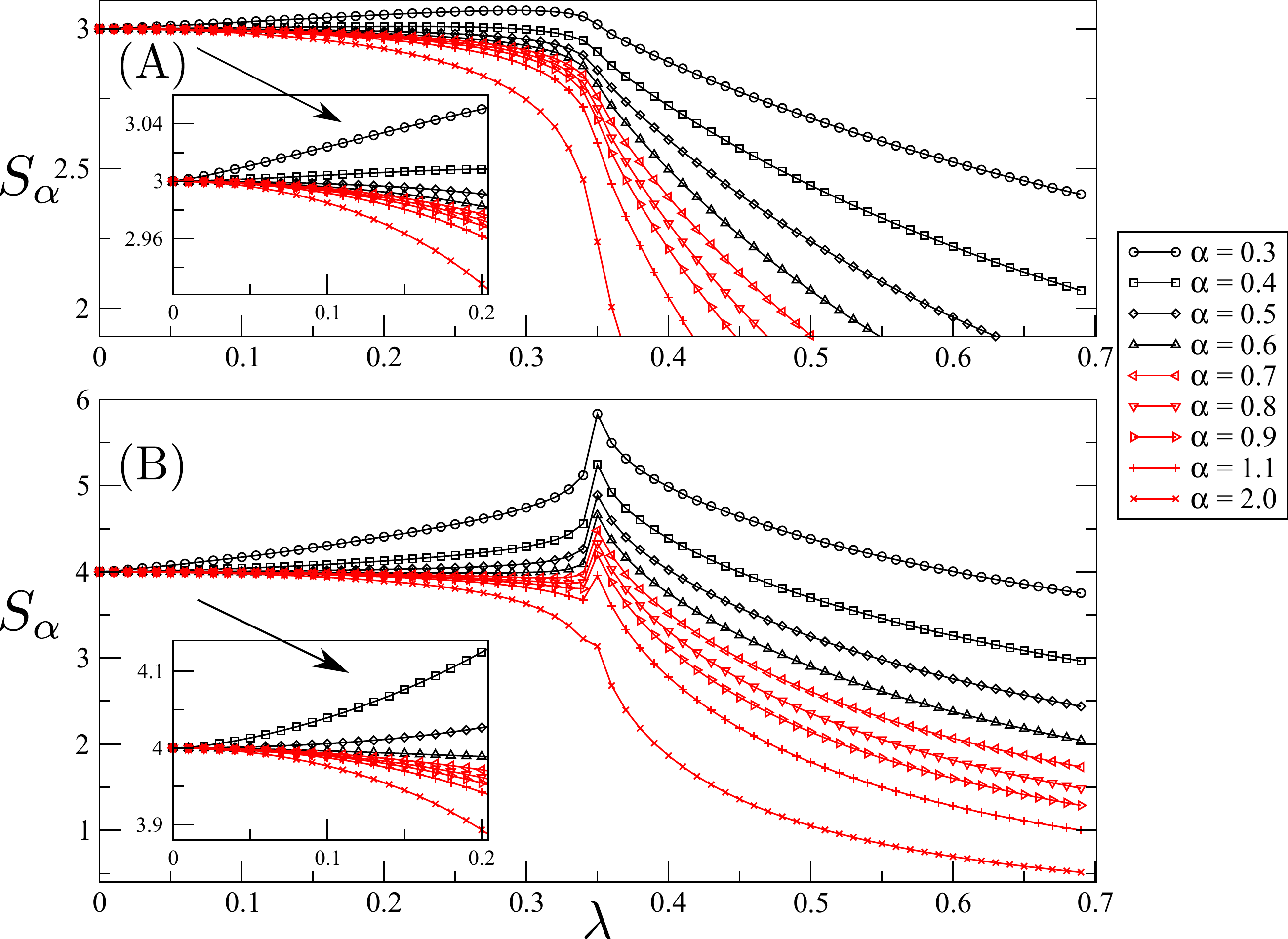}
	\caption {R\'enyi entropies as a function of $\lambda$ for the ground state of $\mathcal H= \mathcal H_{TC} + V_{xz}$ with {$\lambda_x=\lambda$ and $\lambda_z = \lambda/2$}.  Here, $L_y=5$. The reduced system $A$ consists of $A_s$ and $C_\mathrm{h.i.}$ in panels A and B, respectively. As $\lambda$ increases, the correlation length increases. The Schmidt rank $R$ and the low $\alpha<\alpha_0$-R\'enyi entropies increase as well. The value of $\alpha_0$ are $0.4, 0.6$ in panel A and B, respectively. Nevertheless, the total entanglement $S_1$ and all the higher R\'enyi entropies are decreasing with $\xi$. Notice the spike in panel B marking the quantum phase transition to the paramagnetic phase at $\lambda_c \sim 0.35$.}
	\label {p2} 
\end {figure}

In Fig.\ref{3dplot}, we can see the behaviour of the $S_\alpha$ R\'enyi entropies as we span the parameter space $\lambda_x, \lambda_z$ for 
the perturbation $V_{xz}$. {We see clearly that in the topologically 
ordered phase  a splitting of $S_\alpha$'s occur:   $\partial_\lambda S_\alpha \lessgtr 0$ at a given value of $\alpha=\alpha_0$;  we found $\alpha_0\simeq 0.6$} (see caption of \ref{3dplot}). We call this phenomenon  $\alpha$ splitting in the rest of this article. In  the paramagnetic phase all the R\'enyi entropies are monotones with $\lambda$. 
This behaviour is generically independent of the size and shape of the subsystem $A$, as long as $A$ contains some bulk \cite{long}. 
Below we provide an  explanation of the phenomenon. The topologically ordered phase we consider is characterized by  the presence of a 
state (at $\lambda=0$) with  $\xi=0$ and  a flat entanglement spectrum (and an area law)\cite{renyi}. The flat entanglement spectrum implies that small perturbations results in decreasing $S_\alpha$ for $\alpha > \alpha_0$ being $\alpha_0 < 1$, because the distribution becomes less flat {in the most represented eigenvalues in the entanglement spectrum}. In contrast, $S_0$ must increase with $\xi$ as an effect of the perturbation (new degrees of freedom are involved in the entanglement spectrum). 
So the $\alpha$ splitting results from the insertion of a finite $\xi$ in the state evolving from a  state with flat spectrum and zero $\xi$.
We also observe  that such property is shared with the so called $G-$states which include all the topologically ordered quantum double models, and states like the cluster states\cite{hiz3}, and therefore our findings apply to this class of models as well\cite{renyi} (see \cite{kalis} for a discussion of 
the cluster phase diagram).  Here we remark that the splitting effectively distinguishes a class of quantum spin liquids (states with finite correlation length and no local order parameter), which are notoriously very difficult to detect, since one cannot  measure correlation functions of all the possible local observables. To further distinguish non topologically ordered quantum spin liquids like the cluster states from topologically ordered states, we need to measure the degeneracy of the ground state, since the former have a unique ground state while topological states possess a degeneracy protected by topology. 

Moreover, notice that the splitting occurs no matter how we perturb in the plane $\lambda_x, \lambda_z$, and it is therefore a robust property of the phase.  Note again that  in the paramagnetic phase all the $\partial_\lambda S_\alpha$ have the same sign and no splitting ever occurs, which is easily understood from the presence (at very large $\lambda$) of a completely factorized state, see Fig.\ref{p2}.

We remark that the splitting phenomenon  effectively  distinguishes the topologically ordered state from a topologically-trivial ordered state (like a ferromagnet). As  discussed above, the latter states  have typically $S_\alpha$ increasing with $\xi$ and no splitting occurs (see Supplementary Material)\cite{suppl}. 
Summarizing: We can distinguish between the  topological phase and the  paramagnet of (\ref{model}); furthermore, we can distinguish between the topological phase and a symmetry breaking phase.

To corroborate our findings, we resort to exact analysis for suitable perturbations $V(\lambda)$'s. 
We consider two cases {\it i)} $V_h = \lambda_z \sum_{i\in h} \sigma_i^z$, corresponding to placing the  external field $\propto \sigma^z$ only along the horizontal links of the lattice;  {\it ii)} $V(\lambda)= \sum_{s}e^{-\lambda\sum_{i \in s}{\sigma}^z_i}$ leading to the Castelnovo-Chamon model\cite{chamon}. Since these perturbations commute with the plaquette operators of Eq.\ref{model}, the ground state of these models can be written as the superposition of  loop states $\ket{g}$ with amplitudes $\alpha(g)$. A loop state $\ket{g}$ is obtained from the completely polarized state in the $z$ direction, by flipping down all the spins intersected by a loop drawn on the lattice. The corresponding loop  operators $g$ form a group $G$ called the gauge group of these theories.

In the case {\it i)}  the  star operators $\prod_{i \in s } \sigma^x_i$ interact only along the rows of the lattice. The model maps onto the product of arrays of Ising chains by the duality $A_s\to \tau^z_{\mu}$, $\sigma^z_i\to \tau^x_{\mu}\tau^x_{\mu+1}$:  $\mathcal H_{TC} + V_h \mapsto
\mathcal{H}_{ff}=\bigoplus_{i=1}^L(-\lambda\sum_{\mu}\tau^x_{\mu}\tau^x_{\mu+1}-\sum_{\mu}\tau^z_{\mu})$ \cite{Dusuel,Halasz}. The relevant correlators in the variables $\sigma$ can be obtained through the correlators in the dual variables $\tau$ that can be accessed exactly\cite{McCoy}.   In  the following, we sketch a proof that the splitting phenomenon does occur in this model (see \cite{long} for details). 
We consider the star  $A_s=\{ i_1, i_2, i_3, i_4\}$ as subsystem A (see Fig.\ref{cylinder});   $\rho_{A_s}$ is {block diagonal with $2 \times 2$ blocks labeled by $\ket{i_1i_2 i_3 i_4}$ and $A_s\ket{i_1i_2 i_3 i_4}$}. It results that $\rho_A$ has maximum rank unless $\alpha(g) =\alpha(g_1)\alpha(g_2)$, implying there is a zero eigenvalue in each block. 
In the dual picture this  is equivalent to require $\langle \tau_i \tau_j\rangle= \langle \tau_i \rangle \langle \tau_j\rangle$. Such condition holds at  
$\lambda=0$ only, and therefore  $R$ increases at  $\lambda\neq0$}. 
The factorization of the amplitudes also proves that both the $\alpha=1,2$-R\'enyi entropies decrease at small $\lambda$\cite{long}. 

{The case  {\it ii)}  is important to test the argument of the interplay between splitting and increasing of correlation length. This argument implies that
 a perturbation for which $\xi(\lambda)=const$ does not lead to a splitting in the Renyi entropies}. The model of Castelnovo-Chamon  features exactly this, since spin-spin correlation functions $\langle \sigma^x_i \sigma^x_j\rangle$ are vanishing for every value of $\lambda$.
The exact ground state is made of loops with amplitudes $\alpha(g)= e^{-\lambda/2\sum_{i\in s}\sigma^z_i(g)}$, where 
 $\sigma^z_i(g) =  \bra{g} \sigma^z_i\ket{g}$. 
 The  topological phase is($\lambda<\lambda_c\approx 0.44$). 
{A lengthy calculation leads to  
$
S_{\alpha}(\rho_A)=(1-\alpha)^{-1}\log {Z^{-\alpha}(\lambda)}\sum_{g\in G}e^{-\lambda L_g}w^{\alpha-1}(\lambda,g)
$,
{where  $Z= \sum_g e^{-\lambda L_g}$ and $w(\lambda,g):=\sum_{h\in G_A,k\in G_B}e^{-\lambda L_{hgk}}$, $L_m$ is the length of the loop  $m$ of the gauge group $G$; here $G_A$ and $G_B$ are the gauge groups of the subsystems $A$ and $B$ respectively}.
The analysis   of  small and large $\lambda$ expansions reveals that $\partial_{\lambda}S_{\alpha}(\lambda)\le 0$\cite{long}. }As a particular case, $S_0$ is constant for every value of  $\lambda$. Accordingly, for this fine tuned perturbation all R\'enyi entropies  decrease and therefore no splitting is observed. This is consistent with the fact that also in this model the amplitudes $ \alpha(g)$ factorize {as discussed in {\it i)} }.

In this paper, we have shown that topologically ordered states in the paradigmatic class of quantum double models have a peculiar way of responding to perturbations: The R\'enyi entropies of a subsystem display opposite  slopes around $\alpha \simeq 0.6$. The subsystem size can be small independently  of $\xi$. In the paramagnetic phases, in contrast, no such a phenomenon is observed. 
The $\alpha$ splitting is shown to be a robust characteristic of the topological phase, see Fig.\ref{3dplot}. In a quantum information context, we can say that, while paramagnets are always differential locally convertible, the class of topologically ordered ground states we studied here are not.
By measuring $\xi$ together with   the von-Neumann entropy $S_1$ for a small subsystem (it can be shown this is sufficient to show whether $\alpha$ splitting is present), one could tell apart  spin liquid states of the class studied here from symmetry breaking states. This should be contrasted with the standard approach based on topological entropy where the size of the subsystem must be much larger than $\xi$.

Our results may provide the basis for the experimental detection of exotic phases like topologically ordered states or certain quantum spin liquids.  
 In this endeavour, it would be  desirable to investigate these findings in presence of disorder. 

It would be also important to investigate the {behaviour of R\'enyi entropies under perturbations} for 
more general topological states without a flat entanglement spectrum like  fractional quantum Hall liquids and  chiral spin liquids\cite{Papanikolaou,Isakov}, and   for  symmetry protected topological states like one dimensional systems or topological insulators\cite{Cui_symtopo}.  
A relevant  question is  whether the splitting behaviour can shed light on the finite temperature resilience of topological order, and on the very existence of anyonic excitations.

{\em Acknowledgments.---} We thank  J.I. Cirac, R. Fazio, G. Vidal, X.G. Wen for discussions. This work was supported in part by the National Basic Research Program of China Grant 2011CBA00300, 2011CBA00301 the National Natural Science Foundation of China Grant 61073174, 61033001, 61061130540.
Research at Perimeter Institute for Theoretical Physics
is supported in part by the Government of Canada through NSERC and
by the Province of Ontario through MRI.


\begin{references}
\bibitem{caveat} For moderate $\xi$, also S2 splits in the right way, and it is more convenient to measure \cite{abanin, measurerenyi}
\end{references}
\end{document}